\documentclass[twocolumn,final]{article}
\usepackage{preprint}
\usepackage{graphicx}
\usepackage[hidelinks]{hyperref}
\usepackage{tabularx}
\usepackage[normalem]{ulem}
\usepackage{tabularray}
\usepackage{multirow}
\usepackage[bottom]{footmisc}
\usepackage{cite}
\usepackage{amsmath,amssymb,amsfonts}
\usepackage{cleveref}
\usepackage{algorithmic}
\usepackage{textcomp}
\usepackage{flushend}
\usepackage{wrapfig}
\usepackage{subcaption}

\def\BibTeX{{\rm B\kern-.05em{\sc i\kern-.025em b}\kern-.08em
    T\kern-.1667em\lower.7ex\hbox{E}\kern-.125emX}}
\newcommand{\todo}[1]{}

\usepackage{tikz}

\newcommand*\circlednum[1]{\tikz[baseline=(char.base)]{
    \node[shape=circle,draw,very thick,inner sep=1pt] (char) {#1};}}
    
\newcommand*\circlednumsmall[1]{\tikz[baseline=(char.base)]{
    \node[shape=circle,draw,inner sep=1pt] (char) {#1};}}

\makeatletter 
\newcommand{\linebreakand}{%
  \end{@IEEEauthorhalign}
  \hfill\mbox{}\par
  \mbox{}\hfill\begin{@IEEEauthorhalign}
}
\makeatother 

\newcommand{\name}{\textsc{Oxn}}
\AtBeginShipout{\ifnum\value{page}=5\pagecolor{red!20}\fi}

\title{Continuous Observability Assurance in {Cloud-Native Applications}}%

\usepackage{authblk}

\author[1]{Maria C. Borges}
\author[1]{Sebastian Werner} 
\affil[1]{Information Systems Engineering,
Technische Universität Berlin, Germany}

\begin{document}
\twocolumn[\begin{@twocolumnfalse}
\maketitle
\pagestyle{plain}

\begin{abstract}
When faults occur in microservice applications -- as they inevitably do -- developers depend on observability data to quickly identify and diagnose the issue. 
To collect such data, microservices need to be instrumented and the respective infrastructure configured. 
This task is often underestimated and error-prone, typically relying on many ad-hoc decisions.
However, some of these decisions can significantly affect how quickly faults are detected and also impact the cost and performance of the application.

Given its importance, we emphasize the need for a 
method to guide the observability design process. 
In this paper, 
we build on previous work and integrate our observability experiment tool OXN into a novel method for continuous observability assurance. We demonstrate its use and discuss future directions.
\end{abstract}

\vspace{0.5cm}

\end{@twocolumnfalse}]

\section{Introduction}Modern cloud-native applications are typically built as microservices, which are developed by different teams, run on complex cloud infrastructure and depend on a diverse software stack \cite{Bogner_MicroservicesStudy_2019}. In such dynamic and evolving environments, not everything can be tested upfront, so these applications are typically more prone to faults \cite{Zhang_MicroserviceSurvey_ThresholdingHard_2019,Niedermaier_ObservabilityInterviewStudy_2019}. When faults occur, developers rely on observability data to quickly identify and diagnose the issue. To collect such data, microservices need to be instrumented and the respective infrastructure configured. 

Designing the observability of an application is a challenging task. Different frameworks in the stack expose different data, parameter configurations often depend on the specific tools used, and developers frequently add custom instrumentation code. This process is often underestimated and heavily relies on ad-hoc decisions or ``professional intuition" \cite{Zhang_MicroserviceSurvey_ThresholdingHard_2019,Vale_MicroserviceQualityMeasurementIndustrySurvey_2022}. Poor choices can  drive up costs or hinder fault detection, which forces developers into reactive measures, instrumenting and reconfiguring only after problems emerge \cite{Srinivas_MonitoringCoverageMicrosoft_2024}. Consequently, this results in improvised solutions rather than systematically derived observability designs \cite{Niedermaier_ObservabilityInterviewStudy_2019,Srinivas_MonitoringCoverageMicrosoft_2024}.

As cloud-native microservice applications continuously evolve and improve, so should the observability of the application. Decisions regarding observability must undergo constant reevaluation to ensure continuous improvement and adaptability to changing reliability requirements. Addressing these challenges demands the establishment of a structured process to continuously measure the observability of an application.  While most research has typically focused on measuring the cost of observability \cite{Ernst_OfflineTraceGeneration_2021,Reichelt_OverheadOpenTelemetryEtc,Dinga_EnergyEfficiencyOfMonitoring_2023}, our earlier work introduced metrics and tooling to measure observability more comprehensively \cite{Borges_observabilitydecisions_2024,Borges_oxnposter_2024}. 

In this paper, we build on that foundation by integrating our previous work into a method for continuous observability assurance. It provides a systematic approach for observability design by incorporating experiment-based assessments proposed in our previous work into the application development lifecycle. It aligns with SRE principles and guidelines \cite{Google_SREBook_2016} to ensure continuous improvement and adaptability to changing reliability trade-offs.

In the following, we show how our method builds upon related industry practices, present the complementary tooling, introduce the method, demonstrate its use and discuss future directions.\label{sec:intro}
\section{Background: Site Reliability Engineering}Originating at Google, Site Reliability Engineering (SRE) \cite{Google_SREBook_2016} is a discipline that employs various practices across an organization to assure application reliability. Of particular interest to our work are the practices of error budgets and postmortems. 
Error budgeting is a practical approach to manage reliability-related trade-offs. By defining an acceptable threshold for system unreliability based on agreed-upon service-level objectives (SLOs), teams can prioritize efforts, balancing between developing new features or conducting reliability improvements.
Postmortems are detailed analyses conducted after production incidents. They try to document what happened, why it occurred, and how to prevent recurrence. The goal is to learn from failures to improve systems and processes in the future.

\begin{figure}[t]
    \centering
    \includegraphics[width=0.8\linewidth]{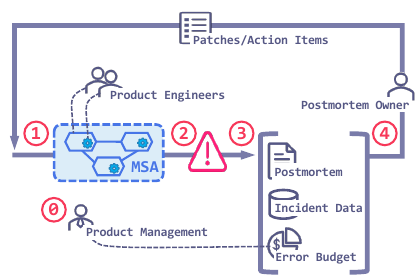}
    \caption{SRE and the Postmortem Cycle} 
    \label{fig:postmortem}
\end{figure}

Here, we describe a formalized process for SRE postmortems, derived from Google’s
guidelines \cite{Google_SREBook_2016}. \Cref{fig:postmortem} illustrates the process.

\noindent\textbf{Step \circlednum{0} -} Product Management defines SLOs for a product, which sets an expectation of how much downtime is tolerated for the application per quarter, the so-called ``error budget". 

\noindent\textbf{Step \circlednum{1} -} Product Engineers develop features as microservices and add observability to the application, including several instrumentation/configuration decisions.

\noindent\textbf{Step \circlednum{2} -} After a period of uptime, the application inevitably runs into a fault and crashes, triggering the incident response process. 

\noindent\textbf{Step \circlednum{3} -} Eventually, the incident response team figures out what the root cause of the problem is. Once the emergency has been mitigated, the team starts the process of writing up a postmortem, which captures the incident details, identifies the root cause, describes the actions taken to resolve it, and outlines preventive actions to adopt in the future. This postmortem document serves as a reference for future incident prevention and response. It is frequently accompanied by incident data, which can be used for retrospective analysis. Lastly, since the application experienced downtime during the incident, the error budget for the quarter is updated accordingly.

\noindent\textbf{Step \circlednum{4} -} One of the authors of the postmortem document becomes the Postmortem Owner and is accountable for follow-up patches and completion of action items. When these are carried out, the Product Engineer team pushes another update in \circlednumsmall{1} and the cycle begins again.

\label{sec:bg}
\section{Background: Observability Experiment Engine}
\begin{figure}[t]
    \centering
    \includegraphics[width=0.95\linewidth]{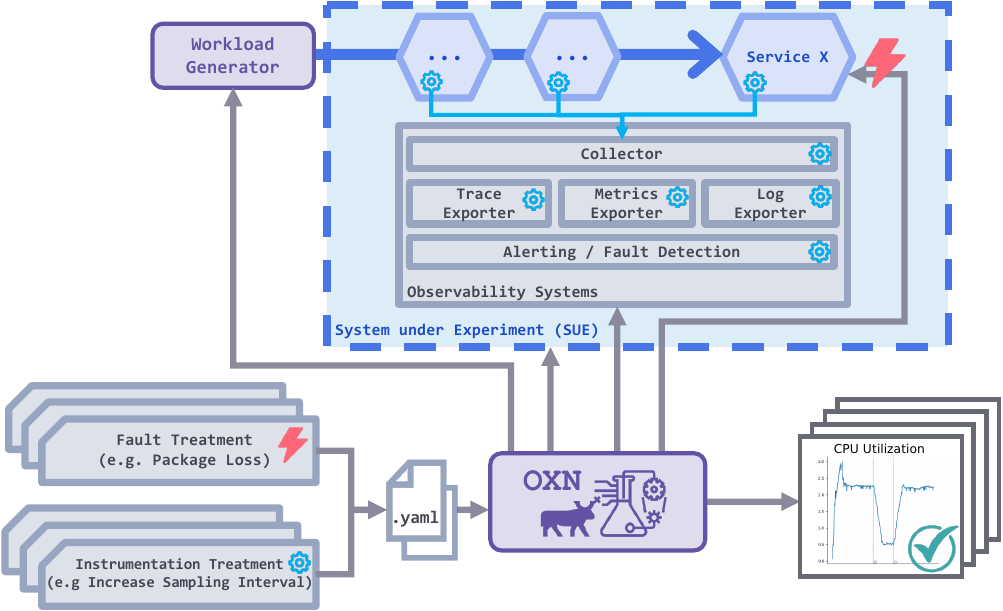}
    \caption{Overview of the tool OXN} 
    \label{fig:oxn}
\end{figure}

To assess observability design trade-offs with concrete evidence and arrive at appropriate instrumentation and configuration, we advocate for conducting observability experiments. 
To facilitate such experimentation, we created OXN~\cite{Borges_oxnposter_2024}, a tool that automates this process.
With OXN, we can evaluate the observability system's effectiveness by injecting faults into the microservice system under experimentation (SUE) and monitoring the configured detection mechanisms, such as alerting thresholds, while also measuring the costs produced by the observability systems.
By observing the system's response to the injected faults, we can assess the effectiveness of the observability system. Figure \ref{fig:oxn} shows an overview of the tool. 

OXN is able to deploy any microservice application specified in a docker-compose file or kubernetes helm chart. After deploying the SUE, OXN starts a workload generator, where load shapes can easily be specified via locust files, a  tool commonly used in industry. For the experiments, OXN introduces treatments, which are controlled changes to the system under experiment. We distinguish between fault treatments and instrumentation treatments. \name{} already provides a core set of treatments out-of-the-box, which cover some basic use-cases and serve as proof-of-concept. Behind these treatments lies a common interface for by practitioners to implement custom treatments, such as custom fault scenarios. OXN also enables the definition and modification of the SUE's instrumentation, for instance, by adjusting sampling frequencies, adding more instrumentation points, or changing altering thresholds.

\label{sec:oxn}
\section{Method} We present a method for continuous observability engineering.
 It provides a systematic approach for observability design decisions by incorporating experiment-based assessments and tooling introduced by OXN~\cite{Borges_oxnposter_2024} into the application development lifecycle. Our method follows SRE principles \cite{Google_SREBook_2016} and hooks into the SRE postmortem cycle, promoting continuous improvement and adaptability to changing reliability needs.

\subsection*{Postmortem Observability Assessments}
Continuous reliability improvements happen when teams learn from past incidents and take preventive measures. Similar to regression testing, we propose turning past faults into  testable incident scenarios. This approach allows teams to spot weaknesses or inefficiencies in the observability machinery and ensures that issues that were once hard to detect are caught early in the future. It also ensures that measures remain in place and hard lessons are not forgotten.

\begin{figure}[t!]

\centering
\begin{subfigure}{\linewidth}
    \hspace{0.5mm}\includegraphics[width=\linewidth]{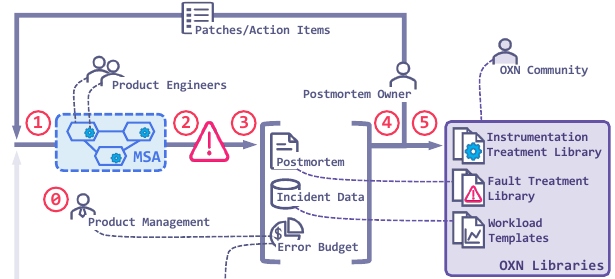}
    \caption{OXN can be extended to support Postmortem Observability Assessments \\}\label{fig:postmortem_oxn}
\end{subfigure}
\hfill
\begin{subfigure}{\linewidth}
    \includegraphics[width=\linewidth]{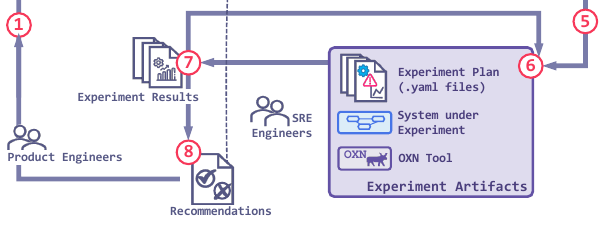}
    \caption{Employing OXN in the application development cycle\\}\label{fig:oxn_experiments}
\end{subfigure}\\[0.5em]
\caption{Method for Continuous Observability Assurance}
\label{fig:method}

\end{figure}

For this purpose, we developed OXN to be extensible and highly customizable to specific application scenarios. Behind each treatment lies a common interface that engineers can leverage to implement custom treatments, for example to simulate a past fault scenario. Here, the open source OXN repository\footnote{https://github.com/nymphbox/oxn} can serve as a platform for engineers to share fault scenarios and instrumentation tuning options across organizations through pull requests. Lastly, a workload generator interface  enables the simulation of incident related user behavior. Thus, we propose the following extension to the Postmortem Cycle (see also \cref{fig:postmortem_oxn}):

\noindent\textbf{Step \circlednum{5} -} The Postmortem Owner extends the OXN fault treatment library to replicate the fault from the recent incident and creates a locust workload for OXN based on the actual incident data. This has two benefits, one the incident becomes testable and thus the patches and system improvement apparent, second, these experiments can be re-run to continuously verify that future modifications don't reintroduce issues, thus, increasing system reliability in the long term.

\subsection*{Continuous Observability Assurance}
In this section, we explain how observability experiments can be employed to continuously improve the observability of microservice applications. \Cref{fig:oxn_experiments} illustrates this process.

\noindent\textbf{Step \circlednum{6} -} The SRE Engineers identify experiments of interest. For each experiment, they create an experiment plan, i.e. .yaml files with a machine readable description of the experiment design, including what to observe, what instrumentation to tweak, what faults to inject, and what load to use. With the rest of the experiment artifacts, they can run experiments.

\noindent\textbf{Step \circlednum{7} -} The SRE Engineers run several experiments and collect results showing how effective different observability configurations are at identifying the faults relevant to the application and how costly they are.

\noindent\textbf{Step \circlednum{8} -} The SRE Engineers identify and evaluate trade-offs. Based on how much downtime can be tolerated in the rest of the quarter (i.e. what is left in the error budget), they might make different recommendations to the Product Engineering team. Both teams negotiate and decide on implementation changes for the next release, completing the feedback loop in the next step (\circlednumsmall{1}).

This method enables a collaborative and systematic design for observability, which can be continuously applied during the development of cloud-native applications. It empowers the broader community to build a reproducible repository of fault scenarios, which can be automaticaly tested using OXN. Furthermore, its integration into the SRE lifecycle allows for effective control over when, how often, and at what cost these observability design decisions need to be updated or refined.\label{sec:architecture}
\section{Preliminary Instantiation}In this section, we present a preliminary demonstration of our method, while acknowledging the need for further 
validation, e.g. through collaboration with industry SRE teams. For the demonstration, we construct an example based on a common fault. 
Specifically, the incident involves a latency spike in a dependent service, a recurring issue documented in both research \cite{Supriyo_microsoftpostmortems_2022} and industry blogs \cite{Gitlab_RedisLatency_2022}. 

As our microservice application, we use the OpenTelemetry Astronomy Shop Demo\footnote{https://github.com/open-telemetry/opentelemetry-demo}. It consists of 20 microservices and is instrumented to collect several metrics and traces across the whole application, making it a suitable candidate to demonstrate the method. We deploy the application on a cloud-based virtual machine (8vCPUs, 32GB Memory).

Since we lack real incident data for experimentation, we simulate a constant load of 50 concurrent users. For instrumentation, our focus is on determining an appropriate sampling rate for tracing data—a configuration option already supported by the OXN tool.
To model a latency spike incident in step 5, we use a network delay fault injection mechanism supported by OXN. This feature allows us to introduce a configurable network delay between two services, which in our case we configure to range between 0-90ms.
As our experiment plan (step 6), we run experiments for 10 minutes each, with 4 minutes allocated for ramp-up, 2 minutes for the fault occurrence, and the final 4 minutes for a return to normal conditions. For more details, consult our delay experiments in the experiment plan\footnote{https://github.com/nymphbox/oxn/tree/main/experiments}.

\begin{figure}[b]
    \centering
    \includegraphics[width=\linewidth]{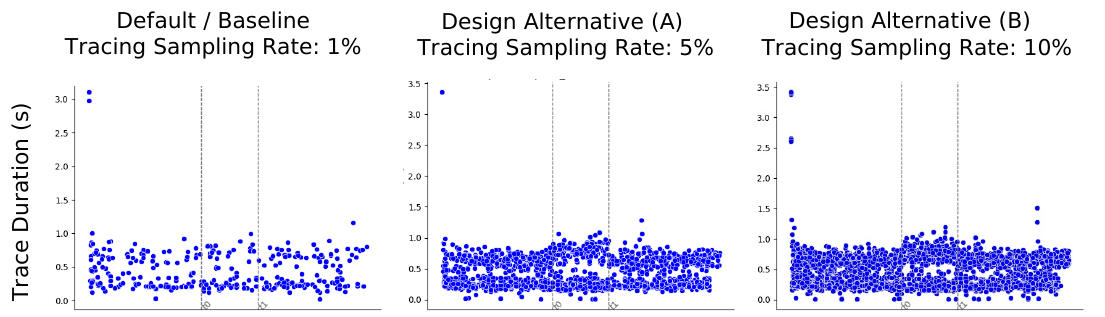}
    \caption{Comparison of Trace Duration Results for Fault Treatment NetworkDelay[0-90ms]} 
    \label{fig:results}
\end{figure}

\begin{table}[]
\caption{Cost of the Design Alternatives}
\centering
\begin{tabularx}{0.65\columnwidth}{l|lll}
         & \footnotesize Baseline                     & \footnotesize A       & \footnotesize B       \\ \hline
\footnotesize CPU time &  \footnotesize191.83s               &  \footnotesize 197.68s & \footnotesize 202.06s \\
\footnotesize Overhead & \multicolumn{1}{r}{-} & \footnotesize +3.05\% & \footnotesize +5.33\%
\end{tabularx}
\label{table:results}
\end{table}

For step 7, we collect results for trace duration across three different tracing sampling rates, 1\%, 5\% and 10\%, as shown in \Cref{fig:results}. The fault is hardly recognizable in the baseline variant, however, it becomes more visible in the two configuration alternatives A and B. The graph shows a slight increase during the fault's duration, an increase also detected by the simple fault classifier implemented in OXN. Table \ref{table:results} shows an overview of the costs for each configuration option, measured as CPU seconds across the relevant services. 

Through the conducted observability experiments, practitioners can see that the fault modeled in our example becomes more visible by increasing the tracing sampling rate. This, of course, comes at a cost. By revisiting step 6, further experiments could explore new metrics to identify the fault at a lower cost. Practitioners would then make recommendations based on the available error budget in step 8. With this demonstration, we hope to introduce the tooling to practitioners and invite them to explore, refine and apply the method in practice.

\section{Related Work}Observability has been evaluated by means of experimentation several times in the past, though these assessments have predominantly focused on the costs associated with observability. For instance, Reichelt et al. \cite{Reichelt_OverheadOpenTelemetryEtc} conducted a concrete benchmark of observability instrumentation. Their study introduced a tool for continuous measurement of overhead of popular instrumentation libraries. 
More recently, Dinga et al. \cite{Dinga_EnergyEfficiencyOfMonitoring_2023} experimentally investigated the energy efficiency of different observability tools. 
In another experiment-based study, Ahmed et al. \cite{Ahmed_EffectivenessOfAPM_2016} explored the effectiveness of four monitoring tools in identifying performance regressions but assumed the default configuration for every tool, therefore ignoring instrumentation and configuration decisions. While all of the three papers investigate an important slice of the observability design landscape, a comprehensive method for continuous observability assurance is still missing. 

Beyond experimentation, Srinivas et al. \cite{Srinivas_MonitoringCoverageMicrosoft_2024} suggest a method for selecting metrics for microservices based on service properties. The paper offers a path to navigate the observability design space, but does not tackle the problem of configuration once the metrics have been selected. Additionally, their approach offers a fixed selection of metrics for each service, which neglects the continuous nature of service development and chances for improvement after production incidents. With the method introduced in our paper, we can continuously validate these metric selections and adjust their configuration for higher fault visibility during incidents.

\section{Conclusion}\label{sec:concl}
In this paper, we propose a method that makes observability design decisions testable and an integral part of continuous software engineering. We outline concrete steps in the application development lifecycle, and align these processes with the SRE postmortem cycle~\cite{Google_SREBook_2016}, incorporating tooling provided by OXN~\cite{Borges_oxnposter_2024}.

Additionally, we argue for a community-driven approach, which can be continuously applied during the development of cloud-native applications. 
Moreover, by rooting our methods in the SRE lifecycle, we benefit from error budgeting, which allows practitioners to balance reliability improvements and feature development, giving a systematic framework for applying OXN effectively.

For future work, we aim to validate and refine this method by applying it to different postmortems in opensource application projects. Besides that, we continue to extend the OXN tool with features to foster broader adoption of experiment-driven observability assurance.

\section*{Acknowledgements}
\footnotesize \noindent Funded by the European Union (TEADAL, 101070186). Views and opinions expressed are, however, those of the author(s) only and do not necessarily reflect those of the European Union. Neither the European Union nor the granting authority can be held responsible for them.

\bibliographystyle{IEEEtran}
\bibliography{refs}

\end{document}